\documentclass[twocolumn%
              ,superscriptaddress%
              ,aps%
              ,prl]{revtex4}

\usepackage{amsmath,amssymb}
\usepackage{graphicx}

\graphicspath{{./Figures/}} % comment out when submitting 

\begin{document}

\title{Quantum-Chaotic Evolution Reproduced from Effective Integrable Trajectories}

\author{Gabriel M.~Lando}
\author{Alfredo M.~Ozorio de Almeida}
\affiliation{Centro Brasileiro de Pesquisas F\'isicas,
Rua Dr.~Xavier Sigaud 150, 22290-180, Rio de Janeiro, R.J., Brazil}
\date{\today}

\begin{abstract}
	Classically integrable approximants are here constructed for a family of predominantly chaotic periodic systems by means of the Baker-Hausdorff-Campbell formula. We compare the evolving wave density and autocorrelation function for the corresponding exact quantum systems using semiclassical approximations based alternatively on the chaotic and on the integrable trajectories. It is found that the latter reproduce the quantum oscillations and provide superior approximations even when the initial coherent state is placed in a broad chaotic region. Time regimes are then accessed in which the propagation based on the system's exact chaotic trajectories breaks down.
\end{abstract}

\maketitle  

\emph{Introduction} -- A fundamental dichotomy between the
quantum and the classical theories in physics is that, while
the first is governed by a linear equation, the latter allows
for much more dynamical complexity due to its general
nonlinearity. Such nonlinearities are a requisite for chaos in
Hamiltonian mechanics, and their absence in the quantum
world indicates that chaos must be somewhat filtered out in
a microscopic description of nature. Research carried out in
the second half of the 20th century has subsequently shown
that even if Schrödinger’s equation forbids chaos, the
quantum mechanics corresponding to classically chaotic
systems can be considered as a field on its own—even
though, strictly speaking, ``there is no quantum chaos, only
quantum chaology'' \cite{Berry1988}. 

An important branch of quantum chaos is dedicated to reproducing quantum dynamics using solely the input extractable from the trajectories of its classical counterpart. This is most often achieved by picking one from a plethora of methods that relate trajectories to quantum objects such as the density of states \cite{Gutwiller1971}, the autocorrelation function \cite{Heller1991}, the Wigner distribution \cite{Ozorio2013} or the wavefunction \cite{VanVleck1928, Herman1984}. These \emph{semiclassical} approximations are usually obtained from asymptotic methods that explore the smallness of $\hbar$ with respect to the typical classical action, so it is expected that $\hbar$ limits the size of the semiclassically relevant phase-space structure. Quantum mechanics should then be immune to the intertwining of classical trajectories, a characteristic of chaotic evolution, in regions with area smaller than $\hbar$ \cite{Berry1979-2,Heller1991}.

There is strong evidence, however, that quantum mechanics can be accurately reproduced by employing classical trajectories even when they are chaotic, despite the ``$\hbar$-area rule" \cite{Heller1991,Lando2019,Lando2019-2}. We here shift direction by investigating the extent to which the trajectories of a specifically tailored integrable system supply a semiclassical approximation for the exact quantum evolution corresponding to a chaotic system -- and for how long. The subject is further enriched by comparing the semiclassical results obtained from the effective (regular) trajectories with the exact (chaotic) ones. Although the substitution of chaotic objects by integrable approximations has been employed in \emph{e.g.}~in chaos assisted tunneling \cite{Brodier2001, Loeck2010} and high harmonic generation \cite{Zagoya2012}, a deeper investigation of this idea has not yet been pursued.

We apply our methods to the propagation of an initial coherent state under the dynamics of the recently introduced ``coserf map" \cite{Lando2019-2}, which is exactly quantizable and has a phase space with mixed regular and chaotic regions. The short, long and very long time-regimes are examined for a kicking strength that renders the system strongly chaotic. The effective integrable system is devised using the Baker-Hausdorff-Campbell formula and its trajectories are obtained using a recently proposed numerical algorithm able to deal with Hamiltonians that are not sums of kinetic and potential terms \cite{Tao2016}. The semiclassical approximations are calculated using the Herman-Kluk propagator, which is very accurate and easily modified to deal with discrete times \cite{Lando2019-2,Maitra2000,Schoendorff1998}. 

\emph{Discrete dynamical systems} --{} Hamiltonians with time dependence of the form 
\begin{equation}
H(q,p;t) = \frac{p^2}{2} + T V(q) \sum_k \delta (t - T k) \, ,  \quad k \in \mathbb{N} \, , \label{eq:timeham}
\end{equation}
where $q$ is the position, $p$ is the momentum and $V$ is a position-dependent potential, present exact solutions to Hamilton's equations and are extensively studied in the context of quantum chaos. The reason for their repeated use is that the corresponding equations of motion are expressed as a discrete map, which can be chaotic even for a single degree of freedom. Here, the sum of delta functions expresses the fact that the potential energy is turned on at times $\tau$, multiples of the \emph{kicking strength} $T$, outside of which the system evolves with constant momentum $p$. The corresponding equations of motion generate stroboscopic maps, \emph{e.g.} the standard map \cite{Chirikov1979}, that split propagation into purely kinetic and purely potential steps. By writing Hamilton's equations for a phase-space point $z = (q,p)$ using Poisson brackets as $dz/dt = \{z, H \}$ we can express the orbits of \eqref{eq:timeham} for a single kick as a composition of two shears generated by two separate Hamiltonians \cite{Lando2019-2}:
\begin{equation}
\begin{cases}
H_1(p) = p^2/2 &\Longrightarrow \,\, U_T^1(\cdot) = \exp \left( - T \left\{ H_1, \cdot \right\} \right) \\[4pt]
H_2(q) = \, V(q)  &\Longrightarrow \,\, U_T^2 (\cdot) = \exp \left( - T \left\{H_2, \cdot \right\} \right) 
\end{cases} \, . \label{eq:sys}
\end{equation}
Using the group property of the solutions above, the final point at $\tau = N T$ for $N$ kicks with kicking strength $T$ is
\begin{equation}
	z_N = U_{T}^N (z_0) = ( U_T^2 \, U_T^1 )^N (z_0) \, . \label{eq:flow}
\end{equation}

Since the flow can be decomposed as successive mappings of the integrable steps in \eqref{eq:flow}, which are exactly quantizable, the corresponding quantum propagation is exact. Quantization for each Hamiltonian evolution in \eqref{eq:flow} is then straightforwardly given by $q \mapsto \hat{q}$, $p \mapsto \hat{p}$, and $\{\, ,\,\} \mapsto [\, ,\, ]/i \hbar$, so that $U_T^j \mapsto \hat{U}_T^j, \, j = 1,\, 2,$ without the need of any ordering considerations. We shall focus on an initial coherent state centered at $z_0 = (q_0,p_0)$:
\begin{equation}
\langle q \vert z_0 \rangle \!=\! (\pi \hbar ) ^{-\frac{1}{4}} \exp \left\{ - (q - q_0)^2/2 + i p_0 (q - q_0)/\hbar \right\} \, , \label{eq:coh} 
\end{equation}
for which the exact quantum evolution in position representation after $N$ kicks with kicking strength $T$ is
\begin{equation}
	 \langle q \vert z_0, N \rangle = \langle q \vert \hat{U}_T^N \vert z_0 \rangle = \langle q \vert ( \hat{U}_T^2 \, \hat{U}_T^1 )^N \vert z_0 \rangle \, . \label{eq:qflow}
\end{equation}

\emph{Effective Hamiltonians} -- Using the Baker-Hausdorff-Campbell formula \cite{Scharf1988} we can approximate the two steps in \eqref{eq:flow} by an effective one:
\begin{equation}
	 e^{-T \{ H_1, \cdot \} } \, e^{-T \{H_2, \cdot \}}  \approx e^{-T \{\mathcal{H}, \cdot \} }\, , \label{eq:comp}
\end{equation}
where, up to third order in $T$, 
\begin{align}
	\mathcal{H} &= H_1 + H_2 + (T/2) \{H_1, H_2 \} + (T^2/12) \left\{ H_1 - \right. \notag \\
	&\left. H_2,  \left\{ H_2, H_1 \right\} \right\} - (T^3/24)\{ H_2, \{H_1 , \{ H_1, H_2 \} \} \} \, . \label{eq:eff}
\end{align}
The \emph{effective Hamiltonian} $\mathcal{H}$ above is time-independent, so its solutions for a period $T$ can be considered as perturbations of the original system for both the classical and quantum cases. Note also that $\mathcal{H}$ cannot be generally expressed as a sum of potential and kinetic energies due to terms proportional to $\{ H_1, H_2 \}$ not vanishing -- a Hamiltonian of this type is known as \emph{non-separable} (even though the system itself is integrable). This implies that solving the equations of motion associated to $\mathcal{H}$, namely 
\begin{equation}
dz/dt = \{ z, \mathcal{H} \} \, , \label{eq:nonsep}
\end{equation}
is best done through the use of special numerical integrators that both preserve the invariants of classical mechanics (such as phase-space areas) and can be applied to non-separable functions. These integrators are called \emph{non-separable symplectic integrators}, and until very recently were limited to algorithms given in terms of computationally expensive implicit functions, being only accurate for short times. Here, however, we are interested in classical propagation for times long enough for chaotic behavior to set in and dominate phase space. We then implement the \emph{explicit} algorithm recently proposed by M.~Tao \cite{Tao2016}, which consists of injecting the system in a larger phase space where its equations of motion are separable, solving them, and projecting the solutions back. We refer to the original article \cite{Tao2016} for error estimates and an accessible exposition of the method. Naturally, depending on the time-regimes of interest, simpler numerical integration algorithms (\emph{e.g.}~Runge-Kutta or Adams-Bashforth) can be used. For times long enough for the system to perform several revolutions around the origin, however, symplectic methods are usually preferred \cite{Yoshida1990}.

\begin{figure}
	\includegraphics[width=\linewidth]{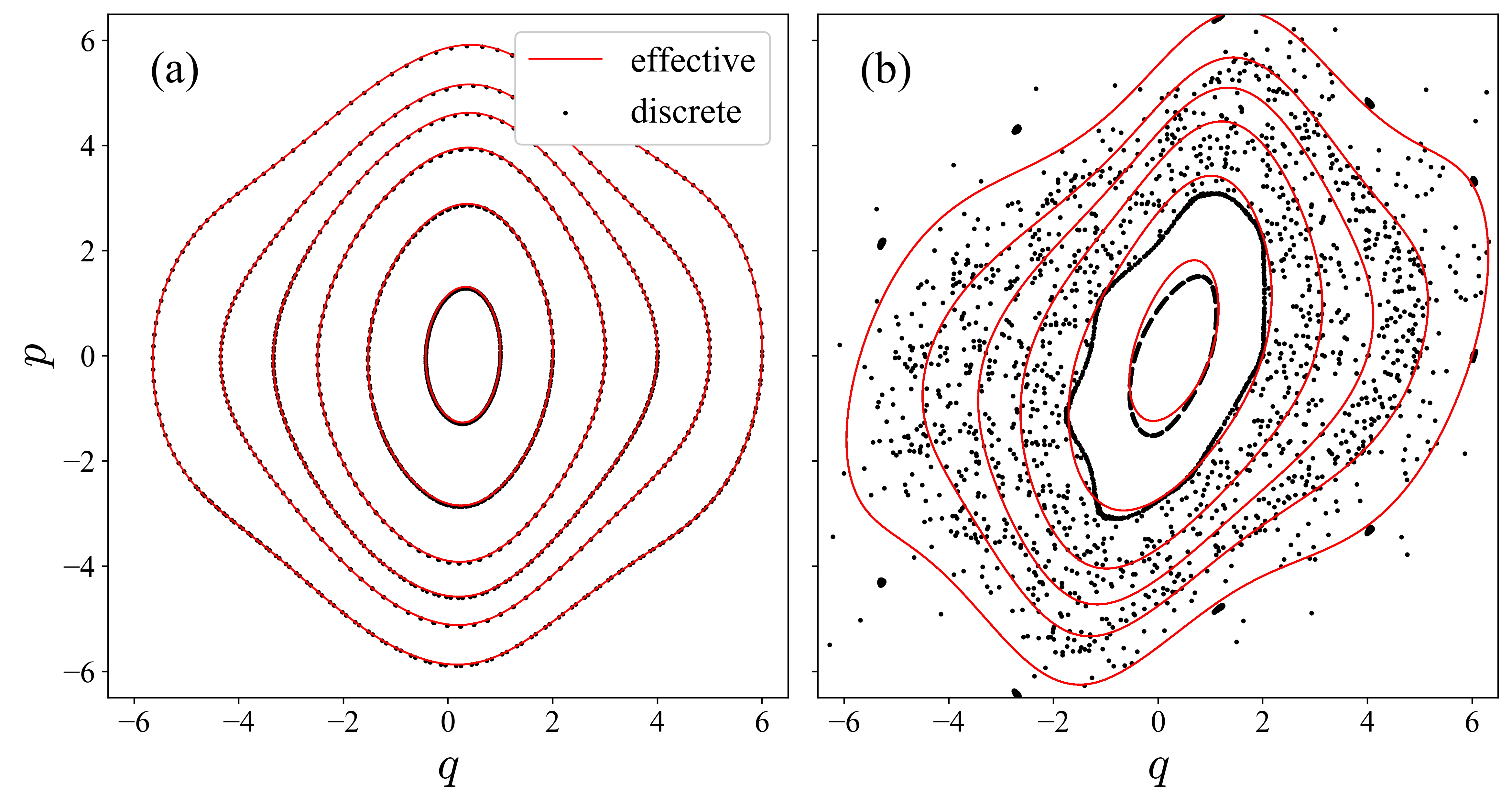}
	\caption{Exact (black dots) and effective (red lines) orbits for the coserf map calculated, respectively, via the iterative map \eqref{eq:flow} and the solutions of \eqref{eq:nonsep} with $\delta = 10^{-2}$. Panel \textbf{(a)} uses $T=0.1$, for which the map displays solely regular trajectories, but for panel \textbf{(b)} we choose $T=0.6$ and a very large chaotic region is presented. Notice how the effective dynamics is simply an interpolation of the exact map for \textbf{(a)},  but for \textbf{(b)} no apparent connection between effective and exact dynamics is seen except for the regular regions near the origin.}
	\label{fig:orbits}
\end{figure}

In Fig.~\ref{fig:orbits} we display some discrete orbits of \eqref{eq:flow} for the \emph{coserf system}, defined by
\begin{equation}
V_{\text{coserf}}(q) = q^2/2 - 2 \cos(q) - \sqrt{\pi} \, \text{erf}(q)/2   \, , \label{eq:erfcosv}
\end{equation}
and their integrable approximations, obtained by applying Tao's method to the effective Hamiltonian $\mathcal{H}$ in \eqref{eq:eff}. All algorithms to integrate Hamilton's equations are discrete, meaning that they have a small iteration step, and the step $\delta$ we used to numerically solve \eqref{eq:nonsep} is small enough for the solutions to look continuous when compared to the discrete dynamics of \eqref{eq:flow}. Notice that even though both $T$ and $\delta$ represent distances between iterations, they are very different in nature: The kicking strength $T$ is seen as a true dynamical parameter that we vary in order to achieve chaos in \eqref{eq:flow}; $\delta$, on the other hand, is just a numerical iteration step that we take as small in order to obtain good accuracy in solving \eqref{eq:nonsep}. We use the simplest version of Tao's algorithm, for which the trajectories obtained from \eqref{eq:nonsep} have errors of $\mathcal{O}(\delta^3)$.

\begin{figure}
	\includegraphics[width=\linewidth]{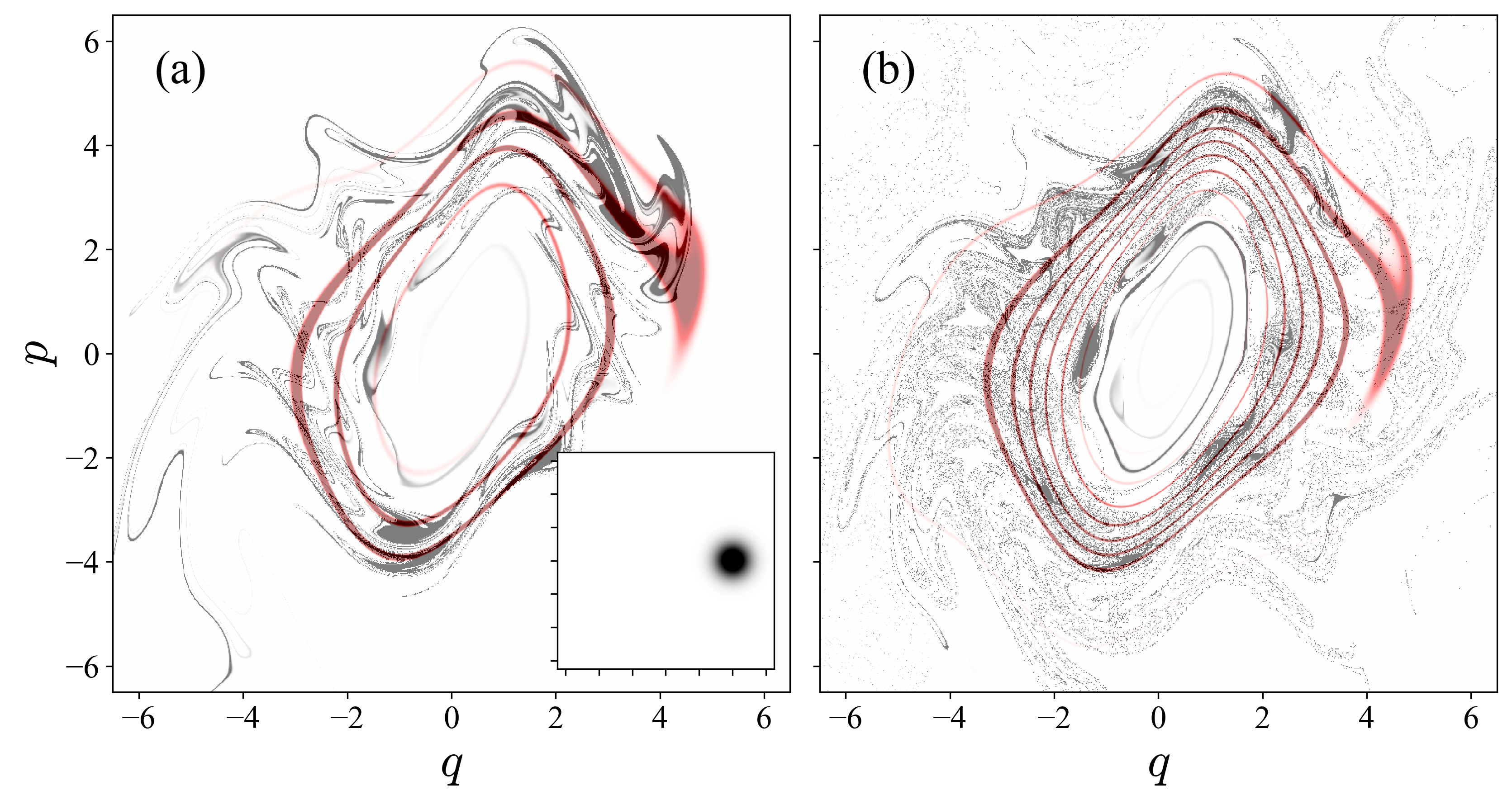}
	\caption{Classical propagation of the Wigner function of a coherent state initially centered at $(q=4,p=0)$. The initial distribution is shown in the inset. The kicking strength is $T=0.6$, as in Fig.~\ref{fig:orbits}\textbf{(b)}, and $\delta = 10^{-2}$ in Tao's algorithm. Chaotic propagation is shown in black, while its effective approximation is superposed in light red. Panel \textbf{(a)} depicts evolution for $N=38$ in \eqref{eq:flow}, while for panel \textbf{(b)} $N=87$.}
	\label{fig:fialments}
\end{figure}

As the trajectories are functions of position and momentum, it is worthwhile to look at how an initial phase-space distribution evolves under both the chaotic and the effective dynamics in order to have a clear picture of their contrast. The obvious choice is the phase-space Gaussian
\begin{equation}
	W(q,p) = \exp \! \left\{-\left[ (q - q_0)^2 + (p - p_0)^2 \right]\!/\hbar \right\}\!/\pi \hbar \, ,
\end{equation}
which can be identified with the Wigner function for the coherent state \eqref{eq:coh} \cite{Ozorio1998}. The evolution of this distribution by classical trajectories corresponds to the approximation of Wigner evolution to lowest order in $\hbar$ \cite{Miller2001,Gro1946,Moyal1949}. The results of both the chaotic and integrable classical evolutions are depicted in Fig.~\ref{fig:fialments}, where it is seen that the initial distribution deforms into a filament that develops ``whorls" and ``tendrils" \cite{Berry1979-2} when exposed to chaotic propagation, but remains completely regular and well-behaved under the effective dynamics. 

\begin{figure}
	\includegraphics[width=\linewidth]{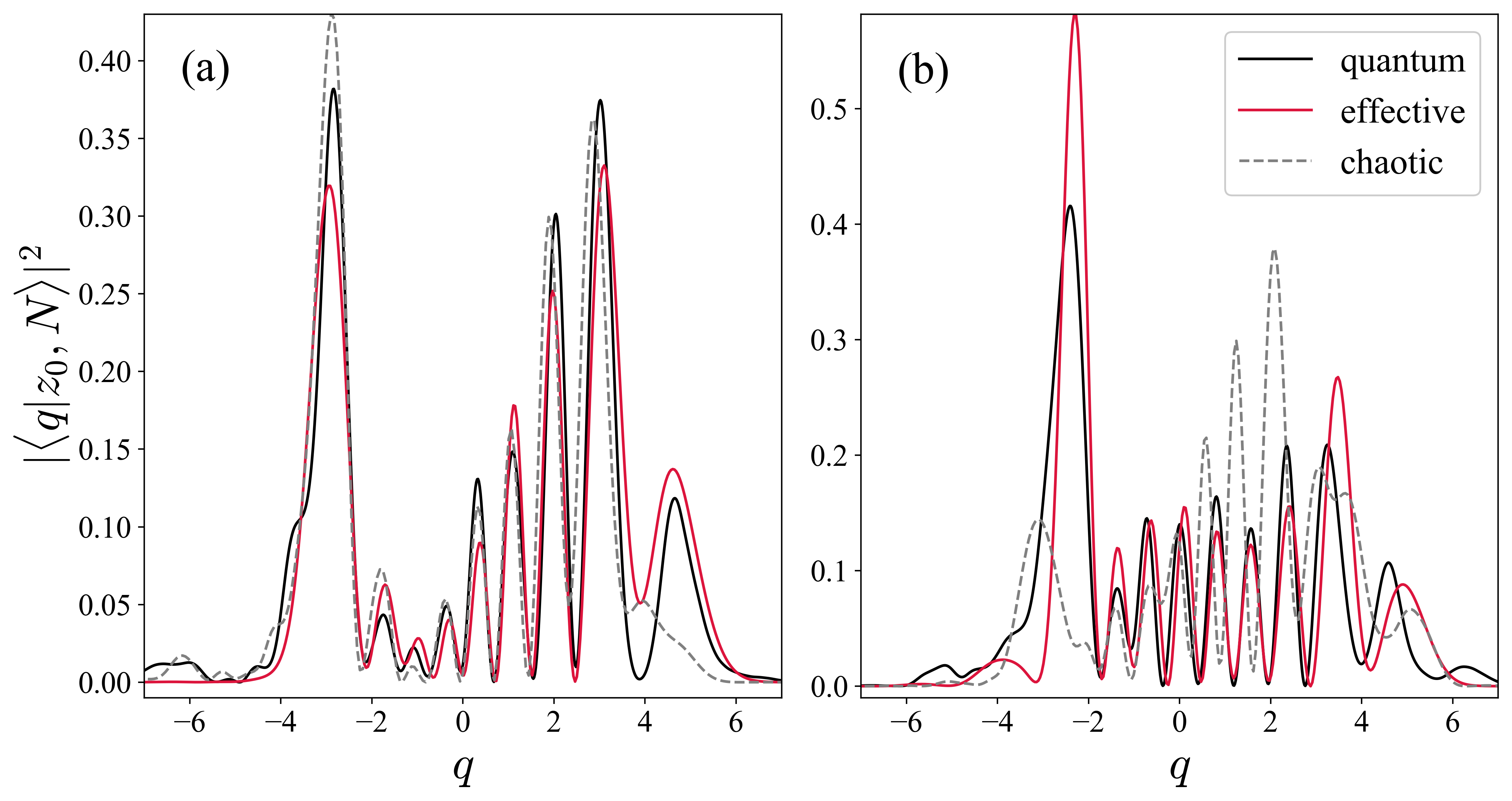}
	\caption{Wave densities for the time-evolution of the coherent state \eqref{eq:coh} obtained via the Herman-Kluk propagation \eqref{eq:qk} and the quantum map \eqref{eq:qflow} for $T=0.6$. The exact quantum result is displayed as a solid black line, while the semiclassical propagations using effective and chaotic trajectories are shown in solid red and dashed gray lines, respectively. We take $\hbar=1$ and, as in Fig.~\ref{fig:fialments}, $\delta = 10^{-2}$, \textbf{(a)} $N=38$ and \textbf{(b)} $N=87$.}
	\label{fig:marginals}
\end{figure}

\emph{The Herman-Kluk propagator} -- Extensively used after its introduction in \cite{Herman1984}, the Herman-Kluk propagator has been adapted to discretized times in several papers \cite{Maitra2000,Schoendorff1998,Lando2019-2}. We express it for $\tau = T N$ as
\begin{equation}
k_{N}(Q', Q) \! = \! \frac{1}{2 \pi \hbar} \int \!\! dz_0 R (z_{N}) \langle Q' \vert z_{N} \rangle \langle z_0 \vert Q \rangle e^{ \frac{i}{\hbar} S (z_{N})} \, , \label{eq:HK}
\end{equation}
where $\langle z_0 \vert Q \rangle$ is the complex conjugate of \eqref{eq:coh} and $\langle Q' \vert z_N \rangle$ is obtained from substituting $(q_0,p_0)$ by $(q_N,p_N)$ in \eqref{eq:coh}. In \eqref{eq:HK}, $\vert Q \rangle $ and $\vert Q' \rangle$ are position eigenstates, $dz_0 = dq_0 \, dp_0$ and 
\begin{align}
R (z_{N}) &= \sqrt{\frac{1}{2} \left[ \frac{\partial p_{N}}{\partial p_0} + \frac{\partial q_{N}}{\partial q_0} + \frac{i}{\hbar} \frac{\partial p_{N}}{\partial q_0} - i \hbar \frac{\partial q_{N}}{\partial p_0} \right]} \label{eq:pre} \\[4pt]
S(z_{N}) &  = T \sum_{k=1}^N \left[ p_k \left( \frac{q_k - q_{k-1}}{T} \right) - H(q_k, p_k) \right] \, .
\end{align}
The square root in \eqref{eq:pre} can and usually does change branch in the complex plane throughout evolution, and it is fundamental to keep track of these changes in order to match the final phases (a procedure known as Maslov tracking \cite{Swenson2011}). The semiclassical approximation for the propagation of a coherent state by the Herman-Kluk method is, therefore, 
\begin{equation}
\langle Q \vert z_0, N \rangle \approx \int dQ' \, k_{N}(Q', Q) \langle Q' \vert z_0 \rangle \, . \label{eq:qk}
\end{equation}
When implementing this formula for the map \eqref{eq:flow} we take $\tau = T N$, and for the effective trajectories that solve \eqref{eq:nonsep} we use $\tau = \delta M$ in Tao's algorithm, where $M$ is chosen such that the final propagation times are the same for both the chaotic and the effective orbits, \emph{i.e.}~$TN = \delta M \Rightarrow M = N(T/\delta)$. These propagation times were already used in Fig.~\ref{fig:fialments}. The semiclassical wave densities $| \langle q | z_0, N \rangle|^2$ for both propagation schemes are plotted against the exact quantum result in Fig.~\ref{fig:marginals} for the same time values as in Fig.~\ref{fig:fialments}. 

As usual in the field, the semiclassical wave densities in Fig.~\ref{fig:marginals} obtained from the chaotic dynamics need to be renormalized in order to have $\int dq \, | \langle q \vert z_0, N \rangle |^2 = 1$, since it is well-known that the wave functions obtained via the Herman-Kluk propagator might lose normalization due to the effect of rapidly separating chaotic orbits in its pre-factor \eqref{eq:pre} \cite{Liberto2016}. The wave function obtained using effective trajectories, however, comes out entirely normalized, since the obstruction to full normalization is due exclusively to chaos. The regularity of the effective trajectories is also responsible for providing very stable results, for which changing grid sizes reflects exclusively on visual resolution; This is in stark contrast with the propagator based on chaotic trajectories, which suffers large deviations depending on the initial grid. Complex procedures to dampen the effect of extreme sensitivity to initial grids in chaotic propagation can be implemented, as in \cite{Tomsovic1993}. A further advantage of the method of effective Hamiltonians is not requiring such artifacts. 

\emph{Discussion} -- As we can see in Fig.~\ref{fig:fialments}, the chaotic propagation is markedly different from its integrable approximation, which interpolates the chaotic regions in phase space as if they were regular (see Fig.~\ref{fig:orbits}). The quantum coserf map, however, has an \emph{exact} classical counterpart, so that it is expected that replacing its true classically chaotic orbits by integrable ones should result in at least some degree of loss with respect to the exact dynamics. In Fig.~\ref{fig:marginals}\textbf{(a)}, quite contrary to intuition, the semiclassical propagator employing the effective trajectories is shown to be as accurate as its chaotic twin; In Fig.~\ref{fig:marginals}\textbf{(b)}, however, we see that it allows for the exploration of time regimes in which the classical distribution propagated using chaotic trajectories has deformed into a stain, and its corresponding semiclassical propagator performs poorly. The effective trajectories, therefore, do not only establish a new connection between quantum mechanics and classical integrability, but also provide a valuable method to reach long times in practical calculations.

\begin{figure}
	\includegraphics[width=\linewidth]{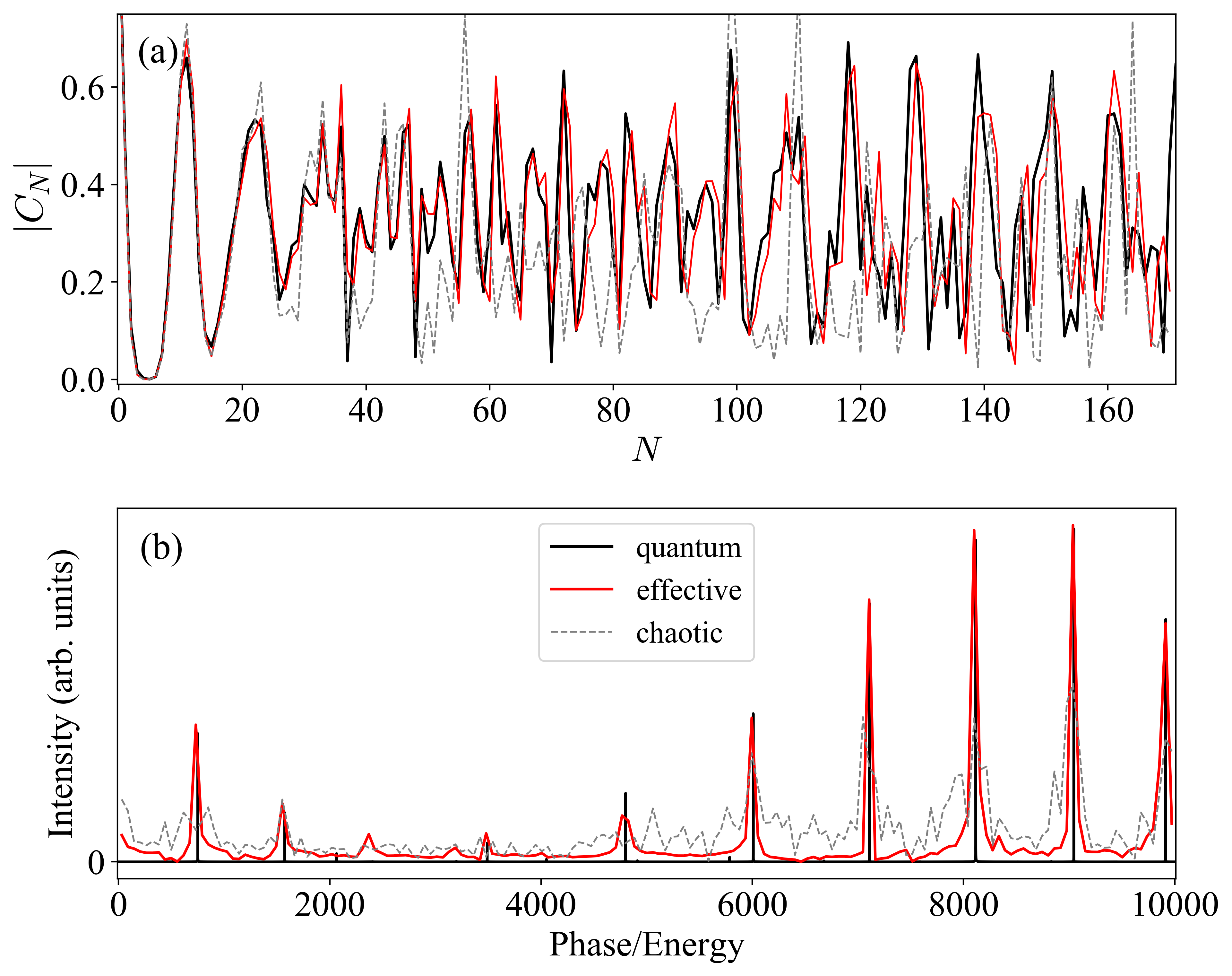}
	\caption{\textbf{(a)} Autocorrelation for the quantum propagation (solid black) and its semiclassical approximations, obtained from the effective (solid red) and the chaotic (dashed gray) trajectories. Although $N$ is discrete, we connect the points for ease of visualization. The parameters are the same as in Fig.~\ref{fig:fialments}, and the Ehrenfest time $\tau_\text{E}$ is near $N = 20$. \textbf{(b)} The discrete Fourier transform of the autocorrelation has peaks at the coserf map's eigenphases, which are well approximated by the eigenenergies of the effective system.}
	\label{fig:result}
\end{figure}

A time threshold exists before which both semiclassical approximation schemes are expected to be equivalent: This is known in the field as the \emph{Ehrenfest time} $\tau_\text{E}$, defined as the moment at which the classical and the quantum autocorrelation functions start to deviate \cite{Lando2019,Schubert2012}. The equivalence in this short-time regime is expected because chaos has not yet impacted classical propagation very strongly. In order to make this discussion more quantitative, in Fig.~\ref{fig:result}\textbf{(a)} we compare the absolute value of the autocorrelation function $|C_N| = |\langle z_0, 0 | z_0, N \rangle |$ for both semiclassical propagation schemes with the exact quantum result. As we can see, the autocorrelation based on effective trajectories fares remarkably well, especially if one considers that $N=170$ corresponds to almost $9 \tau_\text{E}$; The autocorrelation based on chaotic trajectories, however, breaks down near $3\tau_\text{E}$. The approximately 6000 trajectories used to obtain Fig.~\ref{fig:result}\textbf{(a)} were enough when using the effective method, while for chaotic propagation even 35000 trajectories did not provide good results for times longer than $3 \tau_\text{E}$. Worse yet, adding a single trajectory to the grid employed in chaotic propagation drastically changes the final result. As a further means of comparison, in Fig.~\ref{fig:result}\textbf{(b)} we display the discrete Fourier transforms of the autocorrelations in panel \ref{fig:result}\textbf{(a)}, which present intensity peaks at the eigenphases/eigenenergies of a quantum system \cite{Tomsovic1993,Gutwiller1971}. As we can see, the eigenenergies of the effective system accurately resolve even the low-intensity eigenphases of the coserf map, while the chaotic dynamics is seen to add spurious oscillations between the approximate peaks.

The effective propagation, as expected, loses accuracy as we increase the kicking strength, but its failure is generally anteceded by the one of the propagation employing the exact chaotic trajectories. The fact that quantum-chaotic evolution could be better reproduced from an integrable Hamiltonian also indicates that the latter's quantization lies very close to the exact quantum map, since the Herman-Kluk propagator has been shown to be remarkably accurate for integrable systems \cite{Lando2019-2}. It is then expected that more aspects regarding the quantization of classically chaotic systems can also be obtained from chaos-free methods. Although we have used a stroboscopic map due to its visual appeal and exact quantum evolution with which to compare semiclassical results, we remark that there is no obstruction to employing the formalism described here to continuous chaotic systems in higher-dimensional phase spaces. The effective integrable trajectories could be then obtained from \emph{e.g.}~normal forms \cite{Arnold1989,Ozorio1992} or other related methods. The semiclassical propagator used, namely Herman-Kluk's, was chosen due to its implementation ease and demonstrated reliability \cite{Lando2019-2}, but plays no fundamental role and can also be substituted by other methods (such as the one in \cite{Ozorio2013}).

\emph{Conclusion} -- We have shown that the quantum mechanical propagation of a coherent state whose classical analog
has a mixed phase space can be semiclassically approxi
mated very accurately by substituting the original chaotic
trajectories with effective integrable ones. Besides suggesting that chaos might be avoidable in reproducing
quantum evolution, the resulting effective semiclassical
approximation was seen to be even more accurate than
the original one for very long times, presenting itself as a
useful tool to access deep chaotic regimes.

\emph{Acknowledgements} -- We thank R.~O.~Vallejos, F.~S.~Batista and A.~R.~Hern\'andez for stimulating discussions. Partial financial support from CNPq and the National Institute for Science and Technology: Quantum Information is gratefully acknowledged.

\bibliographystyle{apsrev4-1}	 
\bibliography{effective.bib}

\end{document}